# Do the Tellegen particles really exist in electromagnetics?

E.O. Kamenetskii, M. Sigalov, and R. Shavit

Ben-Gurion University of the Negev, Beer Sheva 84105, Israel

July 22, 2008

**Abstract**

In 1948 Tellegen suggested that an assembly of the lined up electric-magnetic dipole twins can construct a new type of an electromagnetic material. Till now, however, the problem of creation of the Tellegen medium is a subject of strong discussions. An elementary symmetry analysis makes questionable an idea of a simple combination of two (electric and magnetic) dipoles to realize local materials with the Tellegen particles as structural elements. In this paper we show that in his search of sources with local junctions of the electrical and magnetic properties one cannot rely on the induced parameters of small electromagnetic scatterers. No near-field electromagnetic structures and classical motion equations for point charges give a physical basis to realize sources with the local junction of the electrical and magnetic properties. We advance a hypothesis that local magnetoelectric (ME) particles should be the physical objects with eigenmode oscillation spectra and non-classical symmetry breaking effects. Our studies convincingly prove this assumption. We show that a quasi-2D ferrite disk with magnetic-dipolar-mode oscillations is characterized by unique symmetry features with topological phases resulting in appearance of the ME properties. The entire ferrite disk can be characterized as a combined system with eigen electric and magnetic moments. The fields near such a particle are distinguished by special symmetry properties.

## I. INTRODUCTION

The question on relations between magnetoelectricity and electromagnetism is a subject of a strong interest and numerous discussions in microwave and optical wave physics and material sciences. An idea about a local magnetoelectric (ME) effect in media goes back to Debye who suggested in 1925 the possible existence of molecules which have a permanent electric dipole moment as well as a permanent magnetic dipole moment [1]. Tellegen considered an assembly of electric-magnetic dipole twins, all of them lined up in the same fashion (either parallel or anti-parallel) [2]. Since 1948, when Tellegen suggested such "glued pairs" as structural elements for composite materials, the electrodynamics of these complex media was a subject of serious theoretical studies (see, e.g. [3 – 5]). Till now, however, the problem of creation of the Tellegen medium is a subject of strong discussions. The questions – How the "glued pairs" of two small (electric and magnetic) dipoles can be realized as local ME sources for electromagnetic waves? and, more generally, Do the Tellegen particles really exist in electromagnetics? – are still open.

The electric polarization is parity-odd and time-reversal-even. At the same time, the magnetization is parity-even and time-reversal-odd [6]. These symmetry relationships make questionable an idea of a simple combination of two (electric and magnetic) small dipoles to realize local ME materials for electromagnetics. If one supposes that he has created a particle with the local cross-polarization effect one, certainly, should demonstrate a special ME field near this particle. It means that using a gedankenexperiment with two quasistatic, electric and magnetic, point probes for the ME near-field characterization, one should observe not only an electrostatic-potential distribution (because of the electric polarization) and not only a magnetostatic-potential distribution (because of the magnetic polarization), one also should observe a special cross-potential term (because of the cross-polarization effect). This fact contradicts to classical electrodynamics. What will be an expression for the Lorentz force acting between these particles? This expression should contain the "electric term", the "magnetic term", and the "ME term". Such an expression is unknown in classical electrodynamics. One cannot consider (classical electrodynamically) a system of two coupled electric and magnetic dipoles as local sources of the ME field and there are no two coupled Laplace equations (for magnetostatic and electrostatic potentials) in the near-field region [6]. In a presupposition that a particle with the near-field cross-polarization effect is really created, one has to show that inside this particle there are internal dynamical motion processes with special symmetry properties.

Such well known "electromagnetic ME scatterers" as small helices [4], $\Omega$-paricles [7] and split-ring-resonators (SRRs) [8] are, in fact, small delay-line sections with distinctive inductive and capacitive regions. In microwave experiments with these particles, no "ME coupling" was shown in the standing-wave systems. For realization of the effect of "ME coupling" in these special-form small scatterers one should have the propagating-wave behavior. Moreover, the "ME response" will be dependable on the type of propagating electromagnetic fields. For the plane, cylindrical, and spherical electromagnetic waves, there will be different "ME responses". The question is still open: What kind of the near-field structure surrounding such an "electromagnetic ME scatterer" should one expect to see? Any quasistatic theories (similar, for example, to the quasistatic Lorentz theory used for artificial dielectrics [9]) are not applicable for such metallic-inclusion composites. In paper [10], authors declared about experimental realization of the ferrite-based Tellegen particle in microwaves. The particle, created as a small ferrite sphere combined with a small piece of a thin metal wire, was placed in a rectangular waveguide cavity. The "ME parameter" of such a particle was estimated via an amplitude of a "cross-polarized wave". It is well known, however, that for a microwave resonator containing enclosed gyrotropic-medium samples, the electromagnetic-field eigenfunctions will be complex, even in the absent of dissipative losses. It means that one does not have the standing-wave fields in spite of the fact that the eigen frequencies of a cavity with gyrotropic-medium samples are real [11]. A microwave resonator with a ferrite inclusion acting in the proximity of the ferromagnetic resonance (FMR) is a nonintegrable system with the time-reversal symmetry breaking effect. The waves reflected from the ferrite boundary are chaotic random waves in a cavity [12 – 14]. Since no eigen oscillations inside a ferrite sphere are observed, the microwave responses will be dependable on a type of the exciting field. So for such a particle no definite and stable polarizability parameters characterizing the ME properties can be found.

The Tellegen proposition was about the particles with permanent coupled electric and magnetic dipole moments. In all the above "ME scatterers" we have the induced effects. It was stated in [15] that so-called Janus particles can be considered as the Tellegen particles with permanent (electric and magnetic) moments. Janus particles are bifacial nanoparticles [16]. They may be separately electrically dipolar or ferromagnetic. The lack of centrosymmetry in synthetic Janus systems may lead to the discovery of novel material properties. Probably, there can be the ME properties. These properties, however, were observed at DC fields and no wave dynamics effects are shown in the studies [15]. On the other hand, the question on the role of symmetry breaking in ME coupling is essential in ME crystals and piezo-composites. There are well known ME materials which are single-phase non-centrosymmetric magnetic crystals or composites that contain a piezoelectric phase [17, 18]. One can change the material parameters by a bias electric field and observe the ME properties for different wave dynamics processes, even at microwaves [17 – 19]. At the same time, natural magnetoelectric crystals and complex ferrite/piezoelectric structures are not particulate composites with ME (Tellegen-type) particles as structural elements. In a proposition of particulate ME composites one may suppose that the unified ME fields originated from a point ME particle (when such a particle is created) will not be the analytically-described classical electromagnetic fields. It is known that in solids effective interactions of charges and spins are often quite different from the fundamental laws of electrodynamics, which may give rise to unusual phenomena. One can expect that the motion equations inside a conjectural local ME particle may lead to non-classical fields with special symmetry properties.

Recently it was found that magnetic-dipolar-mode (MDM) oscillations in a quasi-2D ferrite disk are characterized by the dynamical symmetry breaking effects [20] resulting in the near-field structures with unique topological properties, which are reflected in microwave experiments by specific responses [21 – 25]. Based on analytical and numerical studies, in this paper we show that MDM ferrite-disk particles exhibit special topological effects and can be observed (by the external near fields) as local systems of two, electric and magnetic, *eigen* moments. These ferrite ME particles cannot be considered, however, as classical electromagnetic "ME scatterers".

The paper will begin with Section 2 giving an analysis of different classical aspects in possible realization of local ME particles. Section 3 will be devoted to consideration of special mechanisms of generation of magnetoelectricity by magnetic chirality. ME properties of quasi-2D ferrite disk particles with magnetic-dipolar-mode oscillations will be shown in Section 4. From the electromagnetic point of view, a ferrite ME particle behaves as a non-integrable object. Such an electromagnetically chaotic system can be well modeled, however, by numerical studies based on the commercial numerical electromagnetic-simulation programs. We present numerical verifications for unique ME properties of thin-film ferrite disks. Section 6 is devoted to numerical studies of ME properties of MDM ferrite disks with special geometries. The paper will be concluded by a summary in Section 6 with an outlook to future developments of structures with local ME properties.

## II. SEARCH OF CLASSICAL SOURCES WITH LOCAL JUNCTIONS OF THE ELECTRICAL AND MAGNETIC PROPERTIES

As worthy argumentation, we should forestall our main analysis with necessary discussions on possible classical point sources with local junctions of the electrical and magnetic properties. Presently, there are numerous publications regarding different classical ways of realization of a Tellegen particle. We will give now some basic aspects on local electromagnetic scatterers, near-field structures and motion equations, which clearly prove our standpoints that no particles with the local cross-polarization effect can be realized from a classical point of view.

**A. Local electromagnetic scatterers**

In electromagnetics, local scatterers are systems whose individual dimensions are small compared with a wavelength. It is evident that if a "ME scatterer" exists as a classical object, the fields and radiation of such a conjectural scatterer should be the fields and radiation of a localized oscillating source in classical electrodynamics. In Maxwell equations, the potentials, fields, and radiation can be considered as being originated from a localized system of charges and currents which vary sinusoidally in time:

$$\rho(\vec{r},t) = \rho(\vec{r})e^{i\omega t},$$
$$J(\vec{r},t) = J(\vec{r})e^{i\omega t}.$$

(1)

For a case when the wavelength is much bigger than sizes of a region occupied by charges and currents, the incident fields induce electric and magnetic multipoles that oscillate in definite phase relationship with the incident wave. There are two limit regions: (a) the near (static) zone and (b) the far (radiation) zone. The near fields are quasi-stationary, i.e. they are oscillating harmonically as $e^{i\omega t}$, but otherwise static in character. Since the fields are static in character, no interactions between the electric and magnetic multipoles are presumed. A general solution of Maxwell equations can be represented, for example, as a series for the Hertzian-vector solution [26]. A character of the expansion depends on frequency and geometrical properties of the current distribution. The first term in the expansion describes the field exciting by an oscillating electric dipole, while the second term in the expansion represents the field stipulated by an oscillating magnetic dipole and electric quadrupole [26]. Only the lowest multipoles, usually electric and magnetic dipoles, are important. These induced dipoles can be calculated from static or quasistatic boundary-value problems [6, 26]. The electric dipole moment $\vec{p}^e$ is defined by the electric polarizability and an electric component of an incident field, while the magnetic dipole moment $\vec{p}^m$ is defined by the magnetic polarizability and a magnetic component of an incident field. For example, for a small dielectric (with a dielectric constant $\varepsilon$) sphere of radius $a$ one has [6]

$$\vec{p}^e = \frac{\varepsilon - 1}{\varepsilon + 2} a^3 \vec{E}_{inc}. \tag{2}$$

There is no magnetic dipole moment. For a small perfectly conducting sphere of radius $a$ one has

$$\vec{p}^e = a^3 \vec{E}_{inc}, \qquad \vec{p}^m = -\frac{1}{2} a^3 \vec{B}_{inc}. \tag{3}$$

So a small conducting obstacle exhibits an electric dipole polarization as well as magnetic dipole polarization. One also has the induced electric and magnetic moments for a more complicated case of scattering of electromagnetic fields by a small gyrotropic sphere. In this case, as well, the electric dipole is induced by the electric component of the incident field and the magnetic dipole is induced by the magnetic component of the incident field [27]. From a classical point of view, there is no physical mechanism for *interaction* between electric and magnetic dipoles in the near (static) zone. The total field is a superposition of the partial fields originated from *uncoupled* electric and magnetic dipoles and no cross polarization effects take place. Far away from the scatterer (in the radiation zone) the fields are found to be [28]

$$\vec{E}(\vec{r}, \omega) = \left[ k_0^2 \vec{p}^e + \vec{\nabla}\left( \vec{p}^e \cdot \vec{\nabla} \right) + i k_0 \nabla \times \vec{p}^m \right] \frac{e^{i k_0 |\vec{r} - \vec{r}_0|}}{|\vec{r} - \vec{r}_0|}, \tag{4}$$

$$\vec{H}(\vec{r}, \omega) = \left[ k_0^2 \vec{p}^m + \vec{\nabla}\left( \vec{p}^m \cdot \vec{\nabla} \right) - i k_0 \nabla \times \vec{p}^e \right] \frac{e^{i k_0 |\vec{r} - \vec{r}_0|}}{|\vec{r} - \vec{r}_0|}, \tag{5}$$

where $k_0 = \omega/c$. The role of electric dipole $\vec{p}^e$ in the magnetic field structure as well the role of magnetic dipole $\vec{p}^m$ in the electric field structure become evident only in the far field zone.

The fact that there is no physical mechanism for interaction between electric and magnetic dipoles in the near (static) zone and that only in the far (radiation) zone one can observe the effect of "ME interaction" becomes evident not only for small scatterers with simple geometry but also for small scatterers with complicated geometry. This clearly follows from a classical multipole theory. Multipole expansions in electrodynamics provide a powerful method of characterizing electromagnetic fields [29]. A classical multipole theory describes an effect of "ME coupling" when there is time retardation between the points of the finite-region charge and current distributions and this time retardation is comparable with time retardation between the origin and observation points. In such a case, an expression for the field contains combinations of both magnetic and electric multipole moments [29]. One may obtain the EM-wave phase shift between the points of the finite-region charge and current distributions, $\varphi_1$, comparable with the EM-wave phase shift between the origin and observation points, $\varphi_0$, even for a very small scatterer. To obtain such an effect of "ME coupling" one should make a scatterer in a form of a small $LC$ delay-line section. In the far zone of this scatterer we will observe "ME coupling". This can be explained with help of Fig. 1. Let a characteristic size of a scatterer be $l$ and $l \ll L$, where $L$ is a distance between the origin point and the observation point $P$. Let $k_1$ be the wavenumber of the EM wave propagating in a $LC$ delay line and $k_0$ be the wavenumber of the EM wave in vacuum. In a case when $k_1 \gg k_0$, one may obtain $\varphi_1 = kl \approx \varphi_0 = k_0 L$. All the proposed "electromagnetic ME scatterers" [4, 7, 8] have a typical form of a delay-line section with distinctive inductive and capacitive regions. In a series of experimental papers one can see that the "ME coupling" effect in these particles was observed only in the propagation-wave behavior, without any near-field characterizations [30 – 34]. This fact has a clear explanation. When a small special-form "ME-coupling" scatterer is located in a cavity, both phase shifts $\varphi_1$ and $\varphi_0$ become equal to zero. Thus, no "ME coupling" takes place in the standing-wave systems and the observed special properties of the fields scattered from a small "ME particle" are caused, in any event, by the field retardation effects. No such scatterers with local cross-polarization effects can be presumed from classical electrodynamics. This gives a clear evidence why the multipole theory demonstrates the "ME coupling" in a non-local medium [29]. It follows, for example, that retrieval of the effective constitutive parameters of bianisotropic metamaterials from the measurement of the $S$ parameters [35] should be relied only on the far-field characterization.

A dense composite material means an artificial structure with characteristic sizes of structural elements and distances between them much less than the electromagnetic-wave wavelength. One can realize dense composite dielectric and magnetic materials. There is pure static (quasistatic) electric interaction between neighboring metallic rods in an artificial dielectric [9] and there is pure static (quasistatic) magnetic interaction between neighboring dielectric resonators [36] and SRRs [37] in artificial magnetic materials. Physically, one can create a dense composite based on small delay-line-section "ME particles" as structural elements, but this will not be a composite with local (quasistatic) cross-polarization coupling and so no quasistatic theories (similar, for example, to the quasistatic Lorentz theory used for artificial dielectrics) are applicable for such "ME composites". It is stated in [38] that the separation between the macroscopic and microscopic electromagnetic descriptions is not quite as sharp in bianisotropic media as it is in pure dielectrics due to the fact that the cross-polarization coupling vanishes in the long-wavelength limit. A supposition that one can realize dense particulate composites based on "electromagnetic ME scatterers" raises also a question of boundary conditions. It is well known that for materials with nonlocal properties there should be introduced so called additional boundary conditions (ABCs). The ABCs, being considered as supplementary to standard electromagnetic boundary conditions, are derived from some motion equations in a medium [39, 40]. In a case of nonlocal structures composed by "electromagnetic ME scatterers" no motion equations can be presupposed and so no reliable solutions of the boundary-value problems can be obtained. The known "electromagnetic ME composites" are, in fact, various diffracting structures, which do not have any inherent (different from pure electromagnetic) mechanism of ME interaction. In a case of an "electromagnetic ME particle" one has only imagination of the ME coupling in the far-field region. All the proposed "ME particles" are, in fact, open electrical contours oscillating at a resonance frequency and interacting with an external electromagnetic field as classical radiating systems.

Let us suppose *a priori* that (in spite of the above argumentations) a small electromagnetic particle with the local (quasistatic) cross-polarization effect has been created. When a small electric dipole localizes an electric field and a small magnetic dipole localizes a magnetic field, a small particle with the quasistatic cross-polarization effect should have a special ME field in the near-field region. Our next question concerns the properties of the near-field electromagnetic structures: Can the known near-field electromagnetic structures exhibit themselves as the fields with specific ME properties? We will show that an answer to this question is negative.

**B. Near-field electromagnetic structures**

One of the main aspects attracted the concept of metamaterials was a possibility for the near-field manipulation [41]. In electrodynamics, the near-field EM fields are considered as the evanescent (exponentially decaying) modes. In such a sense, metamaterials can be characterized as structures with tailored electromagnetic response. The importance of phenomena involving evanescent electromagnetic waves has been recognized over the last years. The fact that evanescent waves are more confined than the single tone sinusoid waves and hence contain wider range of spatial frequencies indicates that it may be possible to have no theoretical limit of resolution for the near-field patterns. At present, the near-field manipulation becomes an important factor in new applications, such as near-field microscopy and new material structures. Physically, there can be distinguished different categories of the near-fields [42]. For our purpose, we will consider three types of the near-field EM structures.

(1) Evanescent modes

From a general point of view, the near-field of evanescent modes can be defined as the extension outside a given structure (sample) of the field existing inside this structure (sample). For the Helmhotz equation

$$\varepsilon\mu\, k_0^2 = k_x^2 + k_y^2 + k_z^2, \tag{6}$$

where

$$k_0^2 = \frac{\omega^2}{c^2}, \tag{7}$$

$k_x, k_y, k_z$ are wavenumbers along *x,y,z* axes in a medium, two solutions are possible when $k_x$ and $k_y$ are real quantities. The first solution corresponds to the case

$$\varepsilon\mu\, k_0^2 > k_x^2 + k_y^2. \tag{8}$$

It shows that $k_z$ is a real quantity and one has, as a result, the 3D propagating EM process. The second solution takes place when

$$\varepsilon\mu\, k_0^2 < k_x^2 + k_y^2. \tag{9}$$

So $k_z$ is an imaginary quantity. One has the 2D propagating EM process along *x* and *y* axis and the evanescent-mode fields (the near fields) in *z* direction. The typical examples of such evanescent modes can be demonstrated in the field structures of closed below-cut-off microwave waveguides and open optical waveguides. The importance of evanescent electromagnetic waves was recently realized in connection with emergence of near-field optics microscopy. Taking evanescent waves into account prevents the use of any approximations and requires the detailed solution of the full set of Maxwell equations [43]. As an effective method to study evanescent electromagnetic waves, for solving Maxwell's equations one can use expansion in multipoles where electromagnetic fields are constructed from scalar-wave eigenfunctions of the Helmholtz equation [43, 44].

It is evident that to get extremely big quantities of imaginary $k_z$ one should have extremely big quantities of real $k_x$ and $k_y$. On the way to create a perfect lens based on left-handed metamaterials [41], this fact may become the stronger limitation. Really, the concept of an effective medium cannot be used for a perfect-lens slab of a left-handed metamaterial illuminated by a point source, when evanescent waves have a transverse wavelength of the order of or less than the dimensions of the inclusions or their spacings [45]. Misunderstanding of such a limitation can lead to serious flaws in physics of new material propositions for perfect-lens slabs. As an example, we can refer to paper [46], where a material for a perfect-lens slab is conceived as the dilute mixture of helical inclusions. The fact that these nonlocal inclusions are not in each other near field cast doubts on the vital effect of evanescent waves in such a slab lens.

The near fields of evanescent modes have a pure electromagnetic nature and, evidently, are not related to any specific ME fields.

(2) Quasistatic limit

The quasistatic limit means $|\vec{k}_0| \to 0$ and so $|k_x|, |k_y|, |k_z| \to 0$ as well. In this case, no-wave time-dependable quasistatic fields exist. Such quasistatic electromagnetic fields can be realized only due to local sources: or local-capacitance alternative electric charges with surrounding potential electric fields:

$$\vec{E} = -\nabla \varphi, \tag{10}$$

or local-loop conductive electric currents with surrounding potential magnetic fields:

$$\vec{H} = -\nabla \psi. \tag{11}$$

Spatial distributions of potential $\varphi$ as well as potential $\psi$ are described by the Laplace equation. Examples are the quasistatic fields surrounding tip-structure probes in modern microwave-microscopy devices [47]. Certainly, there is no physical mechanism for possible ME coupling between such local electric and magnetic sources.

(3) Quasistatic oscillations

The symmetry between the electric and magnetic fields is broken in finite temporally dispersive media. In this case, quasistatic oscillations may take place. For quasistatic oscillations

$$k_0 << 1/l, \tag{12}$$

where *l* is the characteristic size of a body. In such oscillations, there are no electromagnetic retardation effects since one neglects or electric, or magnetic displacement currents. The following are some examples of quasistatic oscillations.

(a) Quasistationary EM fields in small metal samples. These fields are described by Maxwell equations with neglect of the electric displacement currents. Inside a sample we have the "heat-conductivity-like" equation for the magnetic field:

$$\nabla^2 \vec{H} = \frac{4\pi\mu\sigma}{c^2} \frac{\partial \vec{H}}{\partial t} . \tag{13}$$

Outside a sample there are the quasistatic-field equations:

$$\nabla \cdot \vec{B} = 0, \quad \nabla \times \vec{H} = 0 . \tag{14}$$

The solutions correspond to imaginary numbers of $k_x, k_y,$ and $k_z$ showing that there are non-stationary decaying fields [39].

(b) Plasmon-oscillation fields are the fields due to collective oscillations of electron density. When one considers a metal or a semiconductor as a composite of positive ions forming a regular lattice and conduction electrons which move freely through this ionic lattice, there can be longitudinal oscillations of the electronic gas – the plasma oscillations. The interface between such a sample and a dielectric may also support charge density oscillations – surface plasmons. In the case of surface plasmon modes, the surface plasmon field decays exponentially away from the interface. For the electrostatic description (one neglects the magnetic displacement current), an electric field is the quasielectrostatic field ($\vec{E} = -\nabla\varphi$). The plasmon oscillations may be characterized by electrostatic wave functions, which are eigenfunctions of the Laplace-like equation. For negative frequency dependent permittivity, one can observe a discrete spectrum of propagating electrostatic modes [48] and electrostatic resonances [49]. Surface plasmons can interact with photons (with the same polarization state) if the momentum and energy conditions are right. There is a link between near-field focusing action and the existence of well-defined surface plasmons [41].

(c) Magnetostatic (MS) oscillations are observed in small temporally dispersive ferromagnet samples [11, 39]. For these quasistationary fields, a magnetostatic description (one neglects the electric displacement current) can be used. So a magnetic field is the quasimagnetostatic field: $\vec{H} = -\nabla\psi$. Inside a ferrite sample one has the Walker equation for MS-potential wave function $\psi$:

$$\nabla \cdot [\vec{\mu}(\omega)\nabla\psi] = 0 , \tag{15}$$

where $\vec{\mu}$ is the permeability tensor. Outside a sample, there is the Laplace equation:

$$\nabla^2 \psi = 0 . \tag{16}$$

For a negative diagonal component of the permeability tensor, the solutions inside a sample may be characterized by real wave numbers for all three dimensions. In this case one can observe a discrete spectrum of propagating magnetostatic modes and magnetostatic resonances [11].

There are no classical physics mechanisms for internal coupling between the electrostatic and magnetostatic oscillations. If one conceives realization of a sample with a simple combination of the plasmon and magnon sources, there will not be a local ME particle surrounded by the unified ME field.

**C. Classical motion equations for point charges**

Could there be any kind of classical motion equations for point charges giving a physical basis to realize sources with the local junction of the electrical and magnetic properties?

As we all know (see e.g. [6, 39]), electromagnetic fields in a medium arise from the microscopic Maxwell equations written for the microscopic electric $\vec{e}$ and magnetic $\vec{h}$ fields, microscopic electric charge density $\rho(\vec{r},t)$ and microscopic electric current density $\rho\vec{v}$:

$$\nabla \times \vec{e} = -\frac{1}{c}\frac{\partial \vec{h}}{\partial t}, \quad \nabla \cdot \vec{e} = 4\pi\rho, \tag{17}$$

$$\nabla \times \vec{h} = \frac{4\pi}{c}\rho\vec{v} + \frac{1}{c}\frac{\partial \vec{e}}{\partial t}, \quad \nabla \cdot \vec{h} = 0. \tag{18}$$

For averaged fields one defines electric polarization $\vec{P}$ and magnetization $\vec{M}$ as

$$\vec{D} \equiv \vec{E} + 4\pi\vec{P}, \quad \vec{H} \equiv \vec{B} - 4\pi\vec{M}. \tag{19}$$

Formally, it can be assumed that a medium is composed with electric and magnetic dipoles. In this case one can write the microscopic Maxwell equations with the microscopic electric charge density $\rho^e(\vec{r},t)$, electric current density $\rho^e\vec{v}^e$, magnetic charge density $\rho^m(\vec{r},t)$ and electric current density $\rho^m\vec{v}^m$ as [50]:

$$\nabla \times \vec{e} = -\frac{4\pi}{c}\rho^m\vec{v}^m - \frac{1}{c}\frac{\partial \vec{h}}{\partial t}, \quad \nabla \cdot \vec{e} = 4\pi\rho^e, \tag{20}$$

$$\nabla \times \vec{h} = \frac{4\pi}{c}\rho^e\vec{v}^e + \frac{1}{c}\frac{\partial \vec{e}}{\partial t}, \quad \nabla \cdot \vec{h} = 4\pi\rho^m. \tag{21}$$

After the averaging procedure, in this case one obtains the standard-form macroscopic Maxwell's equations as well, but for averaged fields one has electric polarization $\vec{P}^e$ and magnetic polarization $\vec{P}^m$:

$$\vec{D} \equiv \vec{E} + 4\pi\vec{P}^e, \quad \vec{B} \equiv \vec{H} + 4\pi\vec{P}^m. \tag{22}$$

An analysis of both the above cases assumes that the motion equations are local equations: the average procedure for microscopic current densities takes place in scales much less than a scale of variation of any macroscopic quantity. At the same time, no ME couplings on the microscopic level can be presupposed in these motion equations. In frames of a classical description, no helical loops (recursion motions) are possible for bound charges and no classical laws describe interaction between linear electric and linear magnetic currents.

Together with electric sources used in the standard electromagnetism one can presume the presence of magnetic sources. An analysis made based on the classical Hamilton principle shows that there cannot be proper equations of motion in which the fields originated from electrically charged particles will exert forces on magnetically charged particles, and vice versa [51]. From the "electric" Langragian density with an electric current source, one derives the standard set of Maxwell's equations with the fields defined as

$$\vec{E} = -\nabla\varphi - \frac{1}{c}\frac{\partial \vec{A}}{\partial t}, \qquad \vec{H} = \nabla \times \vec{A}, \tag{23}$$

where $\varphi$ and $\vec{A}$ are, respectively, the scalar and vector "electric" potentials. At the same time, from the "magnetic" Langragian density with a magnetic current source, one derives Maxwell's equations with the fields defined as

$$\vec{H}' = -\nabla\psi - \frac{1}{c}\frac{\partial \vec{C}}{\partial t}, \quad \vec{E}' = \nabla \times \vec{C}, \tag{24}$$

where $\psi$ and $\vec{C}$ are, respectively, the scalar and vector "magnetic" potentials. The Lorentz forces acting on an electric charge $e$ and a magnetic charge $g$ are defined, respectively, as

$$\vec{F}^e = e\left(\vec{E} + \frac{\vec{v}}{c} \times \vec{H}\right) \tag{25}$$

and

$$\vec{F}^m = g\left(\vec{H}' - \frac{\vec{v}}{c} \times \vec{E}'\right). \tag{26}$$

One can derive the symmetrized set of Maxwell's equations from the summarized, "electric" plus "magnetic", Lagrangian density, but this will not result in the "magnetoelectric" Lorentz forces of the forms

$$\vec{F}^e_{ME} = e\left((\vec{E} + \vec{E}') + \frac{\vec{v}}{c} \times (\vec{H} + \vec{H}')\right) \tag{27}$$

and

$$\vec{F}^m_{ME} = g\left((\vec{H} + \vec{H}') - \frac{\vec{v}}{c} \times (\vec{E} + \vec{E}')\right) \tag{28}$$

giving equations of motion in which the fields associated with electrically charged particles will exert forces on magnetically charged particles, and vice versa [51]. The symmetry properties of magnetic charge and current densities under both spatial inversion and time reversal are opposite to those of electric charge and current densities [6]. So coexistence of electric and magnetic charges must involve some forms of parity violation which do not correspond to symmetries of classical laws. One may expect realizing local ME coupling only when dynamical symmetry breaking occurs.

From the above analysis it follows that in his search of sources with local junctions of the electrical and magnetic properties – the ME particles – one cannot rely on the induced parameters of small electromagnetic scatterers (irrespective of material and geometry of these scatterers). No near-field electromagnetic structures and no classical motion equations for point charges give a physical basis to realize sources with the local junction of the electrical and magnetic properties. It becomes evident that the unified ME fields originated by local ME particles should appear with the near-field symmetry properties distinguishing from that of the electromagnetic fields.

While a local ME particle cannot be realized as a classical scatterer with the induced parameters, it can be created as a small magnetic sample with eigen magnetic oscillations having special symmetry breaking properties. In some magnetic structures one can observe effective interaction of the polarization and magnetization which is described by the laws quite different from the fundamental laws of electrodynamics. These peculiar phenomena will constitute a basis of our search of local ME particles.

**III. MAGNETOELECTRICITY GENERATED BY MAGNETIC CHIRALITY**

The interplay between spin and charge degree of freedom is one of the central issue in solid state physics and the cross correlation between these two degrees of freedom is of particular interests. In solid state structures, one can observe the ME effect from the viewpoint of the electric polarization induced by the applied magnetic fields. This can show a proper way for realization of local ME particles for electromagnetic composite materials.

In some magnetic structures, symmetry arguments are used to construct the coupling between the magnetization vector and the electric polarization vector. For a magnetic crystal an electric polarization can arise in the vicinity of the magnetic inhomogeneity [52, 53]. The polarization has directionality with inversion symmetry breaking. So the polarization can couple to magnetization if the magnetization distribution shows the inversion symmetry breaking properties. This implies that the chiral magnetic ordering can induce an electric polarization. From general symmetry arguments, one has the phenomenological coupling mechanisms between the electric polarization $\vec{p}^e$ and magnetization $\vec{m}$. The invariance upon the time reversal, $t \to -t$, which transforms $\vec{p}^e \to \vec{p}^e$ and $\vec{m} \to -\vec{m}$, requires the ME coupling to be quadratic in $\vec{m}$. The symmetry with respect to the spatial inversion, $\vec{r} \to -\vec{r}$, upon which $\vec{p}^e \to -\vec{p}^e$ and $\vec{m} \to \vec{m}$, is respected when the ME coupling of a uniform polarization to an inhomogeneous magnetization is linear in $\vec{p}^e$ and contains one gradient of $\vec{m}$.

Within a continuum approximation for magnetic properties, the ME interactions responsible for a long-range spatial modulations of magnetization contribute to the Landau-type free energies and are known as Lifshitz invariants. In particular, chiral structures in achiral magnetic systems can arise from the presence of the Lifshitz invariant in the free energy. Without requirements of a special kind of a crystal lattice, the symmetry considerations lead to a ME coupling term in the Landau free energy of the form [53, 54]

$$F_{ME}(\vec{r}) \propto \vec{p}^e \cdot [\vec{m}(\nabla \cdot \vec{m}) - (m \cdot \nabla)\vec{m}]. \tag{29}$$

The term on the right-hand side (RHS) of Eq. (29) is nonzero only if the magnetization $\vec{m}$ breaks chiral symmetry, which is the canonical route towards a strong dependence between $\vec{p}^e$ and $\vec{m}$. On physical grounds, this term can readily be understood. The system sustaining a macroscopic electric polarization $\vec{p}^e$ points out a particular direction in space. Therefore, this polarization can only couple to the magnetization if and only if $\vec{m}$ also has directionality and lacks a center of inversion symmetry. One immediately understands that this occurs when the magnetization is spiraling along some axis. Based on standard vector-algebra transformations, it can be shown from Eq. (29) that the relationship between the electric polarization and the chiral-order magnetization is given by

$$\vec{p}^e \propto \vec{m} \times (\vec{\nabla} \times \vec{m}). \tag{30}$$

It was shown, in particular [53], that magnetic vortices in magnetically soft nanodisks are the inhomogeneous magnetization structures with the induced electric-polarization properties.

Let us consider a magnetically saturated ferrite disk with a normal bias magnetic field directed along *z* axis. For negligibly small magnetic losses, one has from the Landau-Lifshitz magnetization motion equation the linear relation between the harmonically time-varied ($\sim e^{i\omega t}$) local precessing magnetization $\vec{m}$ and RF magnetic field $\vec{H}$ [11]:

$$\vec{m} = \ddot{\chi} \cdot \vec{H}, \tag{31}$$

where

$$\ddot{\chi} = \begin{bmatrix} \chi & i\chi_a & 0 \\ -i\chi_a & \chi & 0 \\ 0 & 0 & 0 \end{bmatrix} \tag{32}$$

is the magnetic susceptibility tensor. Components of tensor $\ddot{\chi}$ are defined as $\chi = \frac{\gamma M_0 \omega_0}{\omega_0^2 - \omega^2}$ and $\chi_a = \frac{\gamma M_0 \omega}{\omega_0^2 - \omega^2}$, where $H_0$ is a bias magnetic field, $M_0$ is the saturation magnetization, $\omega_0 = \gamma H_0$, $\omega_M = \gamma 4\pi M_0$, and $\gamma$ is the gyromagnetic ratio. For such a structure, the vector relation (30) has the following components:

$$(\vec{p}^e)_x = K[\vec{m} \times (\vec{\nabla} \times \vec{m})]_x = K m_y \left( \frac{\partial m_y}{\partial x} - \frac{\partial m_x}{\partial y} \right), \tag{33}$$

$$(\vec{p}^e)_y = K[\vec{m} \times (\vec{\nabla} \times \vec{m})]_y = -K m_x \left( \frac{\partial m_y}{\partial x} - \frac{\partial m_x}{\partial y} \right), \tag{34}$$

$$(\vec{p}^e)_z = K[\vec{m} \times (\vec{\nabla} \times \vec{m})]_z = K \left( m_x \frac{\partial m_x}{\partial z} + m_y \frac{\partial m_y}{\partial z} \right), \tag{35}$$

where *K* is a coefficient of proportionality.

With representation of the in-plane components of a magnetization vector in a ferrite disk as

$$m_x \equiv A(x,y)\,\xi(z)\,e^{i\omega t} \tag{36}$$

and

$$m_y \equiv B(x,y)\,\xi(z)\,e^{i\omega t}, \tag{37}$$

one can rewrite Eqs. (33) – (35) for the real-time electric polarization components as

$$(\vec{p}^e)_x = K \frac{1}{2} \operatorname{Re} \xi \left[ B \left( \frac{\partial B}{\partial x} \right)^* - B \left( \frac{\partial A}{\partial y} \right)^* + \left( B \frac{\partial B}{\partial x} - B \frac{\partial A}{\partial y} \right) e^{2i\omega t} \right], \tag{38}$$

$$(\vec{p}^{\,e})_y = -K\frac{1}{2}\text{Re}\,\xi\left[A\left(\frac{\partial B}{\partial x}\right)^* - A\left(\frac{\partial A}{\partial y}\right)^* + \left(A\frac{\partial B}{\partial x} - A\frac{\partial A}{\partial y}\right)e^{2i\omega t}\right],$$

(39)

$$(\vec{p}^{\,e})_z = K\frac{1}{2}\text{Re}\,\xi\frac{\partial \xi}{\partial z}\left[AA^* + BB^* + (AA + BB)e^{2i\omega t}\right]. \tag{40}$$

Let us represent coefficients $A$ and $B$ as $A = |A|e^{i\delta_A}$ and $B = |B|e^{i\delta_B}$, respectively. It is evident that for a case of a circularly polarized magnetization, when $|A| = |B|$ and $\delta_B - \delta_A = \pm\frac{\pi}{2}$, a component $(\vec{p}^{\,e})_z$ will not be a time-varying quantity and one has precessing of vector $\vec{p}^{\,e}$ around $z$ axis with frequency $2\omega$. In a case of an elliptically polarized magnetization all three components of vector $\vec{p}^{\,e}$ are time-varying.

Let $z = 0$ corresponds to a middle plane of a ferrite disk and let media surrounding a ferrite disk have identical parameters above and below disk planes. If one assumes that function $\xi(z)$, giving a distribution of magnetization along $z$ axis, is a standing-wave function, even or odd with respect to the disk thickness, one obtains from Eq. (40) that for $z > 0$ and $z < 0$ components $(\vec{p}^{\,e})_z$ have different signs. So no total electric flux penetrates through the disk. When, however, media surrounding a ferrite disk above and below the disk planes have different parameters, a total electric flux through the disk may occur.

The above consideration can be used for a case of an external action of an electric field applied to a sample with magnetic chirality. The ME interactions of ferrite samples with external fields can be observed if there are eigen chiral states in a particle. We will show now that a quasi-2D ferrite disk with magnetic-dipolar-mode oscillations is characterized by unique symmetry features with topological phases resulting in appearance of eigen chiral states and ME properties.

## IV. ME PROPERTIES OF QUASI-2D FERRITE DISK PARTICLES WITH MDM OSCILLATIONS

Generally, in classical electromagnetic problem solutions for time-varying fields, there are no differences between the methods of solutions: based on the electric- and magnetic-field representation or based on the potential representation of the Maxwell equations. For the wave processes, in the field representation we solve a system of first-order partial differential equations (for the electric and magnetic vector fields) while in the potential representation we have a smaller number of second-order differential equations (for the scalar electric or vector magnetic potentials). The potentials are introduced as formal quantities for a more convenient way to solve the problem and a set of equations for potentials are equivalent in all respects to the Maxwell equations for fields [6]. The situation can become completely different if one supposes to solve the boundary problem for electromagnetic wave processes in small samples of a strongly temporally dispersive magnetic medium [39]. For such magnetic samples with the magnetostatic resonance behaviors in microwaves, the spectral problem for magnetic-dipolar modes cannot be formally reduced to the complete-set Maxwell-equation representation and one becomes faced with a special role of the MS-potential wave function $\psi(\vec{r},t)$, which acquires a physical meaning in the MDM spectral problem [20, 55] and results in experimental observation of energy shifts of oscillating modes and eigen electric moment properties [21 – 25, 56, 57].

In an assumption of separation of variables for MS-potential wave functions in a quasi-2D ferrite disk, a spectral problem in cylindrical coordinates $z, r, \theta$ is formulated with respect to membrane MS functions (described by coordinates $r, \theta$) with amplitudes dependable on $z$ coordinate. The main features of the spectrum become evident from boundary conditions on a lateral surface of a disk [20]. An orthogonal spectrum of oscillations is obtained when one solves a characteristic equation for MS waves in an axially magnetized ferrite rod:

$$(-\mu)^{\frac{1}{2}} \frac{J'_\nu}{J_\nu} + \frac{K'_\nu}{K_\nu} = 0, \qquad (41)$$

where $J_\nu, J'_\nu, K_\nu,$ and $K'_\nu$ are the values of the Bessel functions of an order $\nu$ and their derivatives (with respect to the argument) on a lateral cylindrical surface ($r = \Re$), together with a characteristic equation for MS waves in a normally magnetized ferrite slab:

$$\tan(\beta h) = -\frac{2\sqrt{-\mu}}{1+\mu}, \qquad (42)$$

where $h$ is a disk thickness and $\beta$ is a propagation constant along $z$ axis. A quantity $\mu$ is a diagonal component of the permeability tensor $\vec{\mu}$. A constant $\nu$ is an integer quantity. In these solutions one uses homogeneous boundary conditions for a dimensionless membrane MS-potential wave function $\tilde{\eta}(r, \theta)$:

$$\mu \left( \frac{\partial \tilde{\eta}}{\partial r} \right)_{r=\Re^-} - \left( \frac{\partial \tilde{\eta}}{\partial r} \right)_{r=\Re^+} = 0. \qquad (43)$$

A boundary condition (43) corresponds to a so-called essential boundary condition [58, 59]. It appears, however, that solutions for function $\tilde{\eta}$, obtained based on Eqs. (41) and (43), do not satisfy conditions of continuity of the magnetic flux density on a lateral surface of a disk. There exist another type of a dimensionless membrane MS-potential wave function $\tilde{\varphi}$ (different from the wave function $\tilde{\eta}$), which satisfies the conditions of continuity of the magnetic flux density on a lateral surface of a disk. For function $\tilde{\varphi}$ one has the following boundary condition on a lateral surface of a ferrite disk:

$$\mu \left( \frac{\partial \tilde{\varphi}}{\partial r} \right)_{r=\Re^-} - \left( \frac{\partial \tilde{\varphi}}{\partial r} \right)_{r=\Re^+} = -i \frac{\mu_a}{\Re} \left( \frac{\partial \tilde{\varphi}}{\partial \theta} \right)_{r=\Re^-}, \qquad (44)$$

where $\mu_a$ is an off-diagonal component of tensor $\vec{\mu}$. The boundary condition (44) is a so-called natural boundary condition [58, 59]. From Eq. (44) it evidently follows that for a given sign of parameter $\mu_a$ there are different functions, $\tilde{\varphi}_+$ and $\tilde{\varphi}_-$ corresponding to positive and negative directions of an angle coordinate when $0 \leq \theta \leq 2\pi$. So function $\tilde{\varphi}$ is not a single-valued function. It changes a sign when $\theta$ is rotated by $2\pi$. At the same time, function $\tilde{\eta}$ is a single-valued function.

For functions $\tilde{\varphi}$ one has path dependent (that is non-integrable) solutions. In such a case, one cannot use separation of variables in solving the boundary problem for a MDM ferrite disk. This takes place due to the time-reversal symmetry breaking effect (gyrotropy) on a surface of a ferrite disk. It is well known that because of gyrotropy, electromagnetic waves incident on a ferrite-dielectric interface has reflection symmetry breaking [13, 60]. This effect takes place also for magnetostatic waves propagating in a normally magnetized ferrite film and incident on the film edge [61]. Fig. 2 illustrates this situation. As a result of reflection on the film edge at point *A*, magnetostatic-wave rays $1 \to A \to 1'$ and $1' \to A \to 1$ shown in Figs. 2 (a) and 2(b), respectively, acquire different phases. In a ferrite disk, this results in appearance of azimuthally running waves. Let after a rotation around a disk, an azimuthally running magnetostatic wave acquires a phase $\Phi_1$ while for an opposite direction of a bias magnetic field such a phase is $\Phi_2$. It is evident that $|\Phi_1| = |\Phi_2| \equiv \Phi$ and there should be $\Phi_1 + \Phi_2 = 2\pi p$ or $\Phi = p\pi$. Quantities *p* are odd integers. This follows from the time-reversal symmetry breaking effect. A system comes back to its initial state after a full $2\pi$ rotation. But this $2\pi$ rotation can be reached if both partial rotating processes, with phases $\Phi_1$ and $\Phi_2$, are involved. So minimal $p = 1$ and, generally, quantities *p* are odd integers. It means that for a given direction of a bias magnetic field, function $\tilde{\varphi}$ behaves as a double-valued function.

Such a non-integrable problem has two different analytical solutions for the MDM spectral problem. One of the analytical solutions is based on the concept of the path-dependent (or topological) phase factor for orthogonal MS-potential wave functions. In this case, one uses Bessel functions of integer order $\nu$ as basis functions for the representation of orthogonal functions $\tilde{\eta}$ with an introduction of a special phase factor on a lateral border of a ferrite disk. On a lateral border of a ferrite disk one has the following correspondence between double-valued functions $\tilde{\varphi}$ and single-valued functions $\tilde{\eta}$ [20]:

$$(\tilde{\varphi}_\pm)_{r=\Re^-} = \delta_\pm (\tilde{\eta})_{r=\Re^-}, \tag{45}$$

where

$$\delta_\pm \equiv f_\pm e^{-iq_\pm \theta} \tag{46}$$

is a double-valued function. The azimuth number $q_\pm$ is equal to $\pm \frac{1}{2}$ and for amplitudes we have $f_+ = -f_-$ and $|f_\pm| = 1$. Function $\delta_\pm$ changes its sign when $\theta$ is rotated by $2\pi$ so that $e^{-iq_\pm 2\pi} = -1$. One obtains the energy-eigenstate spectrum of MDM oscillations with topological phases accumulated by the boundary wave function $\delta$. The topological effects become apparent through the integral fluxes of the pseudo-electric fields [20]. There should be the positive and negative fluxes corresponding to counterclockwise and clockwise edge-function chiral rotation. The different-sign fluxes are inequivalent to avoid cancellation. Every MDM in a thin ferrite disk is characterized by a certain energy eigenstate and two different-sign fluxes of the pseudo-electric fields which are energetically degenerate. The spectral theory developed based on orthogonal singlevalued membrane functions $\tilde{\eta}$ and topological magnetic currents, shows the ME effect from a viewpoint of the Berry phase connection [20].

Another type of an analytical solution for the MDM spectral problem is to introduce an additional phase factor so that functions $\tilde{\varphi}$ are represented as "rotating Bessel functions" with an integer azimuth number. This additional phase factor appears when one considers function $\tilde{\varphi}$ as an "in-plane" projection of a helical MS wave. Based on this approach one obtains a picture of the fields of oscillating MDMs derived from non-orthogonal functions $\tilde{\varphi}$. An initial stage of studies of helical MS waves in a ferrite disk is given in Refs. [62]. In a case of rotating magnetic fields in a normally magnetized ferrite disk the clockwise and counterclockwise rotations are strongly different. One has to distinguish the field rotating in the same direction as the precessing spins rotate and the field rotating in an opposite direction with respect to the spins. It is evident that energetically preferable is the resonance interaction when the magnetic field rotates in the same direction as the precessing spins.

The first-type MDM spectral problem solutions (based on orthogonal MS-potential wave functions with topological phase factors) we will conventionally call as the *G*-modes while the second-type MDM spectral problem solutions (based on "rotating Bessel functions") we will conventionally call as the *L*-modes. This arises from the fact that *G*-modes are described by the second-order differential operator $\hat{G}$, while *L*-modes are described by the first-order differential-matrix operator $\hat{L}$ [20, 55]. In a case of *G*-modes, loop topological magnetic currents lead to appearance of eigen electric (anapole) moments. This clearly shows (both in theory [20] and experiments [24, 25]) the ME properties of a MDM ferrite disk. The *G*-mode ferrite particle has special symmetry breaking properties and is not a classical scatterer with the induced parameters. In this paper, we will mainly use rotating Bessel functions (*L*-modes) for demonstration of the ME effect in a ferrite particle. In such a representation, the ME properties of a ferrite disk can be analyzed based on the Lifshitz invariant discussed in Section III of the paper. Moreover, for *L*-modes the structures of the fields inside a ferrite disk and in the quasistatic (near-field) region surrounding a ferrite disk can be numerically modeled based on the HFSS-program studies [63] with a very good correspondence between the numerical and analytical results.

Taking into account Eq. (42) we write a solution for function $\widetilde{\varphi}$ in a form:

$$\widetilde{\varphi} \propto J_\ell \left( \frac{\beta r}{\sqrt{-\mu}} \right) e^{-j\ell\theta}, \tag{47}$$

where $\ell$ is a positive integer quantity. For this type of a solution, a characteristic equation (44) is written as:

$$(-\mu)^{\frac{1}{2}} \frac{J'_\ell}{J_\ell} + \frac{K'_\ell}{K_\ell} - \frac{\mu_a \ell}{|\beta|\Re} = 0. \tag{48}$$

When a magnetic field is varied such that every point on a disk surface describes a closed loop, then the system should, at the end of this excursion, return to its original state. This will give a phase factor $e^{-i\ell\theta}$ for each eigenstate. For a very thin disk one can use separation of variables. For a certain magnetic-dipolar mode one has for MS-potential wave function [20, 55, 59]:

$$\psi = C\xi(z)\widetilde{\varphi}(r,\theta), \tag{49}$$

where $\xi(z)$ is an amplitude factor, $C$ is a dimensional coefficient. Inside a ferrite disk ($r \leq \Re$, $0 \leq z \leq h$) one has:

$$\psi(r,\theta,z) = C_\ell J_\ell \left( \frac{\beta r}{\sqrt{-\mu}} \right) \left( \cos\beta z + \frac{1}{\sqrt{-\mu}} \sin\beta z \right) e^{-j\ell\theta}. \tag{50}$$

This function satisfies characteristic equations (42) and (48). Based on such a MS-potential function one defines the magnetic field ($\vec{H} = -\vec{\nabla}\psi$) inside a ferrite disk as the azimuth propagating wave:

$$H_r(r,\theta,z,t) = C_\ell \frac{\beta}{\sqrt{-\mu}} J'_\ell\left(\frac{\beta r}{\sqrt{-\mu}}\right)\left(\cos\beta z + \frac{1}{\sqrt{-\mu}}\sin\beta z\right) e^{-i\ell\theta} e^{i\omega t}, \tag{51}$$

$$H_\theta(r,\theta,z,t) = C_\ell \frac{-i\theta}{r} J_\ell\left(\frac{\beta r}{\sqrt{-\mu}}\right)\left(\cos\beta z + \frac{1}{\sqrt{-\mu}}\sin\beta z\right) e^{-i\ell\theta} e^{i\omega t}, \tag{52}$$

$$H_z(r,\theta,z) = C_\ell \beta J_\ell\left(\frac{\beta r}{\sqrt{-\mu}}\right)\left(-\sin\beta z + \frac{1}{\sqrt{-\mu}}\cos\beta z\right) e^{-i\ell\theta} e^{i\omega t}. \tag{53}$$

With use of Eqs. (51) – (53) we calculated the magnetic field distributions for a normally magnetized ferrite disk. A ferrite disk has the following material parameters: saturation magnetization is $4\pi M_s = 1880$ G and the linewidth is $\Delta H = 0.8$ Oe. Disk diameter is $D = 3$ mm and thickness $t = 0.05$ mm. A disk is normally magnetized by a bias magnetic field $H_0 = 4900$ Oe. We analyzed the MDMs which are classified by numbers $n$ – the numbers of zeros in the Bessel function, corresponding to different radial variations – for the order parameter $\ell = +1$. We considered the field distributions for first two modes ($n = 1, 2$). The MDM resonance frequencies are obtained from solutions of Eqs. (42) and (48). In Fig. 3 we show a gallery of the analytically derived in-plane magnetic field distributions on the upper plane of a ferrite disk for the 1st MDM ($f = 8.548$ GHz) at different time phases. The magnetic field distributions for the 2nd MDM ($f = 8.667$ GHz) at some time phases are shown in Fig. 4. For the known mode magnetic fields we can calculate the vector $\vec{m} \times (\vec{\nabla} \times \vec{m})$. Space orientations of vector $\vec{m} \times (\vec{\nabla} \times \vec{m})$ on the upper and lower planes of a ferrite disk at different time phases $\omega t$ are shown in Fig. 5 for the 1st MDM ($f = 8.548$ GHz). It is very interesting to find that electric polarization vectors [ $\vec{p} \propto \vec{m} \times (\vec{\nabla} \times \vec{m})$ ] exhibit precessing behaviors These precessing motions are in the same directions for the upper and lower halves of a disk. We also have to note that whereas the magnetization precession occurs with frequency $\omega$, the electric polarization precession is with double frequency $2\omega$. The $z$ components of the electric polarization vectors are oppositely directed on the upper and lower planes of a disk.

The MDMs in a thin ferrite disk can be well studied based on the HFSS numerical electromagnetic-simulation program [63]. Fig. 6 (a) shows the HFSS model of a short-wall rectangular waveguide section with a normally magnetized ferrite disk. The spectral characteristics for a thin ferrite disk shown in Figs. 6 (b) and 6 (c) give a very good correspondence between the absorption peak positions obtained from the numerical simulation and the calculated peak positions for the *G*- and *L*-MDMs (for the order parameter $\ell = +1$). The HFSS-program magnetic field distributions in a ferrite disk shown in Figs. 7 and 8, respectively, for the 1st ($f = 8.52$ GHz) and 2nd ($f = 8.66$ GHz) resonance states at different time phases are in a clear correspondence with the analytically derived magnetic field distributions for the *L*-MDMs (see Figs. 3, 4). Based on the HFSS model we can study also the electric field distributions inside a ferrite disk. The electric field distributions for the 1st ($f = 8.52$ GHz) and 2nd ($f = 8.66$ GHz) resonance states at different time phases are shown, respectively, in Figs. 9 and 10.

An analytical study of the electric field distributions for MDMs appears as a very complicated problem. We can, however, use a simple analytical model. The assumptions used in this analytical model are not well defined. Nevertheless, the model shows a very good correspondence with the numerical results of the electric field pictures. Formally, one can suppose that for the monochromatic MS-wave process there exists a curl electric field $\vec{E}$ defined by the Faraday-Maxwell law. One can represent the electric field as follows:

$$\vec{\nabla} \times \vec{E} = -\frac{i}{c} \omega \vec{B} = -\frac{i}{c} \omega \vec{H} - \frac{i}{c} 4\pi \omega \vec{m},$$

(54)

where $\vec{m}$ is RF magnetization. With the $\vec{\nabla} \times$ differential operation for the left-hand and right-hand sides of Eq. (54) and taking into account that $\vec{\nabla} \times \vec{H} = 0$, one obtains:

$$\nabla^2 \vec{E} = \frac{i}{c} 4\pi \omega \vec{\nabla} \times \vec{m} \ . \tag{55}$$

Here we used the relation $\nabla \cdot \vec{E} = 0$ which is relevant if one assumes that the permittivity of a disk is described by a scalar. The electric field can be formally represented as being originated from an effective electric current:

$$\nabla^2 \vec{E} = i\omega \frac{4\pi}{c^2} \vec{j}^{(e)} \ , \tag{56}$$

where

$$\vec{j}^{(e)} \equiv c \vec{\nabla} \times \vec{m} . \tag{57}$$

With use of Eqs. (31), (32) and (51) – (53) we can obtain the components of magnetization $\vec{m}$ and then derive the components of vector $\vec{\nabla} \times \vec{m}$. The components of an effective electric current are the following:

$$j_r^{(e)}(r,\theta,z,t) = icC_\ell \left( \frac{\chi_a \beta^2}{\sqrt{-\mu}} J'_\ell\!\left(\frac{\beta r}{\sqrt{-\mu}}\right) + \frac{\beta \chi \ell}{r} J_\ell\!\left(\frac{\beta r}{\sqrt{-\mu}}\right) \right)\!\left( \sin \beta z - \frac{1}{\sqrt{-\mu}} \cos \beta z \right) e^{-i\ell\theta} e^{i\omega t}, \tag{58}$$

$$j_\theta^{(e)}(r,\theta,z,t) = cC_\ell \left( \frac{\chi \beta^2}{\sqrt{-\mu}} J'_\ell\!\left(\frac{\beta r}{\sqrt{-\mu}}\right) + \frac{\beta \chi_a \ell}{r} J_\ell\!\left(\frac{\beta r}{\sqrt{-\mu}}\right) \right)\!\left( \sin \beta z - \frac{1}{\sqrt{-\mu}} \cos \beta z \right) e^{-i\ell\theta} e^{i\omega t}, \tag{59}$$

$$j_z^{(e)}(r,\theta,z,t) = -icC_\ell \frac{1}{4\pi} \frac{\mu_a}{\mu} J_\ell\!\left(\frac{\beta r}{\sqrt{-\mu}}\right)\!\left( \cos \beta z + \frac{1}{\sqrt{-\mu}} \sin \beta z \right) e^{-i\ell\theta} e^{i\omega t}. \tag{60}$$

The in-plane effective-electric-current distributions on the upper plane of a ferrite disk for the 1st ($f = 8.548$ GHz) and 2nd ($f = 8.667$ GHz) MDMs at different time phases are shown, respectively, in Figs. 11 and 12. As it follows from Eq. (56), the electric field should be 90° shifted with respect to the effective electric current. So from the effective-electric-current distributions one can obtain qualitative pictures for the electric field distributions. There is a very good correspondence of such pictures of the electric fields with the electric field distributions obtained from numerical studies (see Figs. 9 and 10).

From the field structures of *L*-MDMs (as well as from the HFSS-program field structures) it becomes evident that an entire ferrite disk does not exhibit any electric-moment properties. It, however, shows the magnetic-moment behavior. One can see that the in-plane electric fields on the upper and lower planes of a ferrite disk are in opposite directions at any time phase. Since the disk thickness is much, much less than the free-space electromagnetic wavelength, the disk can be clearly replaced by a sheet with effective linear magnetic currents. A surface density of the effective magnetic current is expressed as

$$\vec{n} \times (\vec{E}_{upper} - \vec{E}_{lower}) = -\frac{4\pi}{c} \vec{i}^{\,m}, \qquad (61)$$

where $\vec{E}_{upper}$ and $\vec{E}_{lower}$ are, respectively, in-plane electric fields on the upper and lower planes of a ferrite disk and $\vec{n}$ is a normal to a disk plane directed along a bias magnetic field. Evidently, $\vec{E}_{upper} = -\vec{E}_{lower}$. Following the electric-field pictures, one can conclude that there are rotating linear surface magnetic currents. Non-zero current-line divergence of such magnetic currents gives equivalent magnetic charges. As a result of this equivalent representation, one has an evidence for an in-plane rotating magnetic dipoles for the entire ferrite disk with MDM oscillations. Due to such in-plane rotating magnetic dipoles one has excitation of the *L*-MDMs in a quasi-2D ferrite disk shown in well-known experiments [56, 57].

Nevertheless, microwave experiments [24] give also an evidence for the MDM excitation by RF electric fields oriented normally to a ferrite disk. This excitation takes place because of the electric (anapole) moments of the *G*-MDMs. The mechanisms of an interaction of the MDM anapole moments with the external RF electric fields are discussed in Ref. [20, 25]. Since the spectra of *G*- and *L*-MDMs are almost degenerate [see the peak positions of *G*- and *L*-MDMs in Fig. 6 (c)], the ME properties of a ferrite disk are experimentally observed as the *G*-mode electric (anapole) moment normal to a ferrite disk together with the *L*-mode magnetic moment rotating in the disk plane. Fig. 13 shows a MDM ferrite disk as a ME particle.

## V. ME PROPERTIES OF MDM FERRITE DISKS WITH SPECIAL GEOMETRIES

We consider now a MDM ferrite disk with special geometry properties. There will be a ferrite disk with a dielectric loading and a ferrite disk with a wire surface metallization. Presently, for these structures we do not have analytical models and so we will analyze the HFSS-program numerical simulation results. It will be shown that in this case a MDM ferrite disk has ME properties which are well verified by experimental results.

As we showed above, when media surrounding a ferrite disk have identical parameters above and below disk planes, no electric-moment properties for *L*-modes for an entire ferrite disk take place. When, however, media surrounding a ferrite disk above and below the disk planes have different parameters, distribution of the electric polarization vectors [$\vec{p} \propto \vec{m} \times (\vec{\nabla} \times \vec{m})$] may lead to a total electric flux through the sample. In Fig. 14 we show the HFSS-program electric field distributions for the 1$^{st}$ resonance state for a sample composed as a thin ferrite disk (with the same parameters as in the above studies) loaded with a dielectric disk with a dielectric constant $\varepsilon_r = 15$. One clearly sees that in this case *topological* electric charges appear and, as a result, one has an in-plane rotating electric dipole. It is interesting to note that a rotating electric dipole is in the same orientation as an in-plane rotating magnetic dipole.

Fig. 15 shows the electric field distributions for the 1$^{st}$ resonance state for a sample composed as a thin ferrite disk with a wire surface metallization. In this case one has an in-plane non-rotating electric and in-plane rotating magnetic dipole. A concept of such a ME particle was put forth in 1996 [64]. Then microwave experiments clearly verified the ME properties in such a structure [21 – 23]. Our present results give pictures for the field distributions in these ME particles.

## VI. CONCLUSION

Understanding a fundamental mechanism of a "junction" of electricity and magnetism in a point source underlies comprehensive solution of a problem of unification of electromagnetism and magnetoelectricity. From our analysis it follows that in his search of sources with local cross-polarization properties one cannot rely on the induced parameters of small electromagnetic scatterers (irrespective of material and geometry of these scatterers). No near-field electromagnetic structures and no classical motion equations for point charges give a physical basis to realize sources with the local magnetoelectricity. In the near-field region with local sources, an electromagnetic field falls into an electric *and* a magnetic fields.

It becomes evident that the unified ME fields originated by conjectural local ME particles should appear with the symmetry properties distinguishing from that of the electromagnetic fields. While a local ME particle cannot be realized as a classical scatterer with the induced parameters, it can be created as a small magnetic sample with eigen magnetic oscillations having special symmetry breaking properties. In some magnetic structures one can observe effective interaction of the polarization and magnetization which is described by the laws quite different from the fundamental laws of electrodynamics. These peculiar phenomena, being a consequence of broken time-reversal and space-inversion symmetries, may constitute a basis for search of local ME particles.

In thin-film ferrite disks with the MDM vortex structures one has the chiral states of magnetization which may provide us with a strong dependence between electric polarization and magnetization. Because of the spectral characteristics with energy eigenstates, the MDM ferrite disks are the particles with strong dynamic localization of the ME fields. It turns out that a MS-potential wave function stabilizes two-dimensional states (the membrane states). At the same time, the quasi-2D MDM ferrite disk can be considered as a "chiral magnetic domain". The chiral-magnetic-ordering dynamics has the electric polarization properties. For given magnetization texture in a ferrite disk, the in-plane distribution of the electric polarization will have azimuth variations. It is evident that a disk will be characterized by such invariant as the electric flux only if the azimuth variation of the electric polarization is described by a non-singlevalued function.

The MDM-ferrite ME fields are characterized by the topological-phase effects. There are not the Tellegen-particle fields which are conceptualized as the electric and magnetic fields of a system with strong quasistatic (dipole-type) localizations. We dispute the physical realizability of Tellegen particles as classical ME particles with cross-polarization effects. The questions raised in this paper are very important in a view of the present strong interest in electromagnetic artificial materials with ME properties. The properties of ME fields shown in this paper give very new insights into a problem of local microwave ME composites.

**Figure captions**

Fig. 1. Effect of "ME coupling": a small electromagnetic scatterer in a form of a *LC* delay-line section has phase shifts $\varphi_1 = kl \approx \varphi_0 = k_0 L$.

Fig. 2. Magnetostatic waves propagating in a normally magnetized ferrite film and incident on the film edge. As a result of reflection on the film edge at point *A*, magnetostatic-wave rays $1 \to A \to 1'$ (a) and $1' \to A \to 1$ (b) acquire different phases.

Fig. 3. A gallery of the analytically derived in-plane magnetic field distributions on the upper plane of a ferrite disk for the 1st magnetic-dipolar mode (*f* = 8.548 GHz) at different time phases. (A qualitative picture).

Fig. 4. A gallery of the analytically derived in-plane magnetic field distributions on the upper plane of a ferrite disk for the 2nd magnetic-dipolar mode (*f* = 8.667 GHz) at some time phases. (A qualitative picture).

Fig. 5. Space orientations of vector $\vec{m} \times (\vec{\nabla} \times \vec{m})$ on the upper (a) and lower (b) planes of a ferrite disk at some time phases for the 1st MDM (*f* = 8.548 GHz). (A qualitative picture).

Fig. 6. Spectral characteristics for a thin ferrite disk. (a) The HFSS model of a short-wall rectangular waveguide section with a normally magnetized ferrite disk; (b) absorption peak positions obtained from the numerical simulation; (c) the calculated peak positions for the *G*-MDM and *L*-MDM (for the order parameter $\ell = +1$).

Fig. 7. A perspective view for the numerically modeled magnetic field distributions on the upper plane of a ferrite disk for the for the 1st resonance state (*f* = 8.52 GHz) at different time phases.

Fig. 8. A top view for the numerically modeled magnetic field distributions on the upper plane of a ferrite disk for the for the 2nd resonance state (*f* = 8.66 GHz) at different time phases.

Fig. 9. A top view for the numerically modeled electric field distributions on the upper plane of a ferrite disk for the for the 1$^{st}$ resonance state ($f$ = 8.52 GHz) at different time phases.

Fig. 10. A top view for the numerically modeled electric field distributions on the upper plane of a ferrite disk for the for the 2$^{nd}$ resonance state ($f$ = 8.66 GHz) at different time phases.

Fig. 11. A gallery of the analytically derived in-plane effective-electric-current distributions on the upper plane of a ferrite disk for the 1$^{st}$ magnetic-dipolar mode ($f$ = 8.548 GHz) at different time phases. (A qualitative picture).

Fig. 12. A gallery of the analytically derived in-plane effective-electric-current distributions on the upper plane of a ferrite disk for the 2$^{nd}$ magnetic-dipolar mode ($f$ = 8.667 GHz) at some time phases. (A qualitative picture).

Fig. 13. The ME-particle model of a MDM ferrite disk. Since the spectra of $G$- and $L$-MDMs are almost degenerate the ME properties of a ferrite disk are observed as the $G$-mode electric (anapole) moment $\vec{a}^e$ normal to a ferrite disk together with the $L$-mode magnetic moment $\vec{p}^m$ rotating in the disk plane.

Fig. 14. The electric field distributions for the 1$^{st}$ resonance state for a sample composed as a thin ferrite disk loaded with a dielectric disk. One has an in-plane rotating electric dipole with topological electric charges. A rotating electric dipole is in the same orientation as an in-plane rotating magnetic dipole.

Fig. 15. The electric field distributions for the 1$^{st}$ resonance state for a sample composed as a thin ferrite disk with a wire surface metallization. One has an in-plane non-rotating electric dipole and an in-plane rotating magnetic dipole.

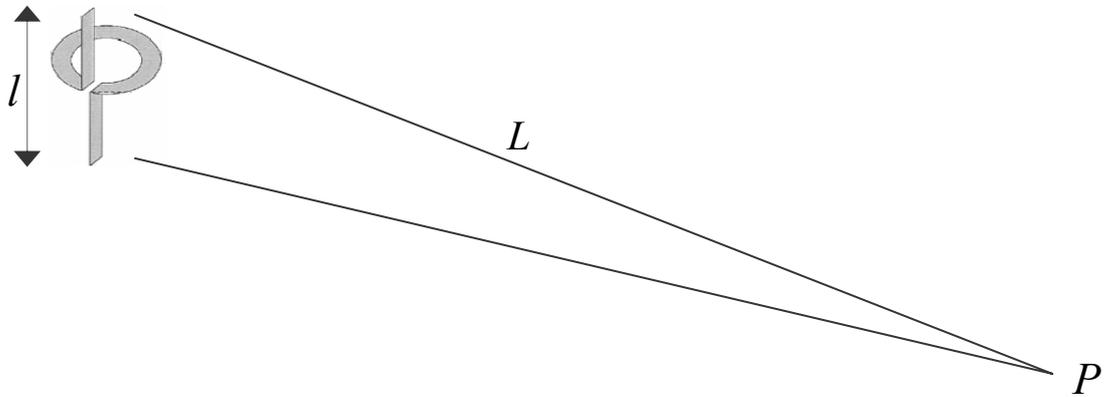

Fig. 1. Effect of "ME coupling": a small electromagnetic scatterer in a form of a *LC* delay-line section has phase shifts $\varphi_1 = kl \approx \varphi_0 = k_0 L$.

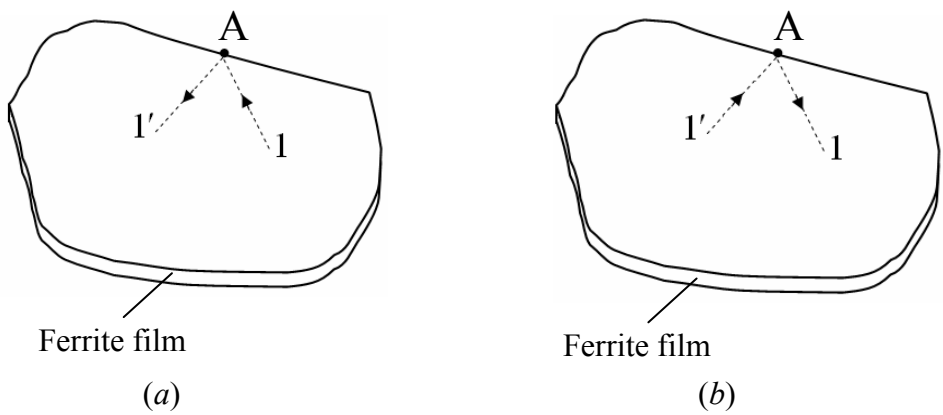

Fig. 2. Magnetostatic waves propagating in a normally magnetized ferrite film and incident on the film edge. As a result of reflection on the film edge at point *A*, magnetostatic-wave rays $1 \to A \to 1'$ (a) and $1' \to A \to 1$ (b) acquire different phases.

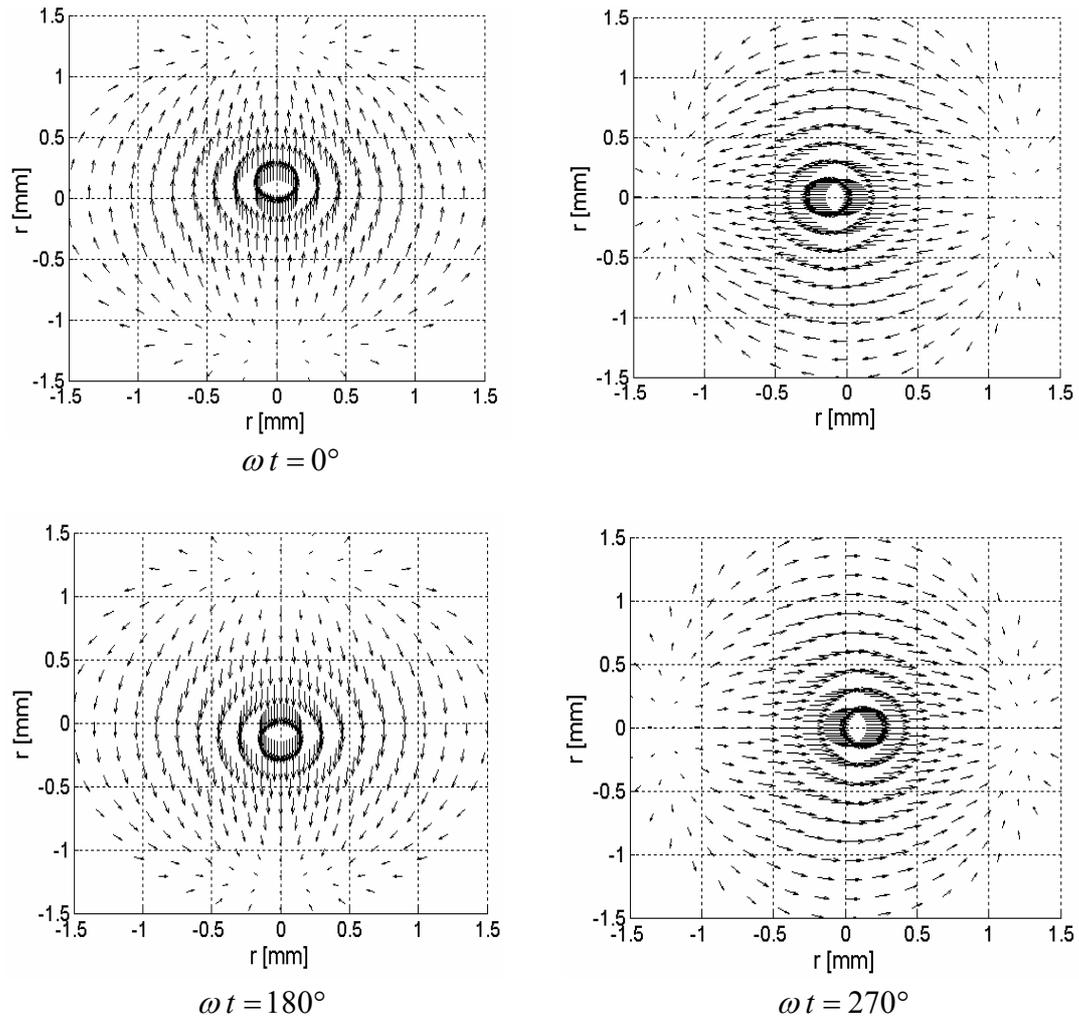

Fig. 3. A gallery of the analytically derived in-plane magnetic field distributions on the upper plane of a ferrite disk for the 1$^{st}$ magnetic-dipolar mode ($f$ = 8.548 GHz) at different time phases. (A qualitative picture).

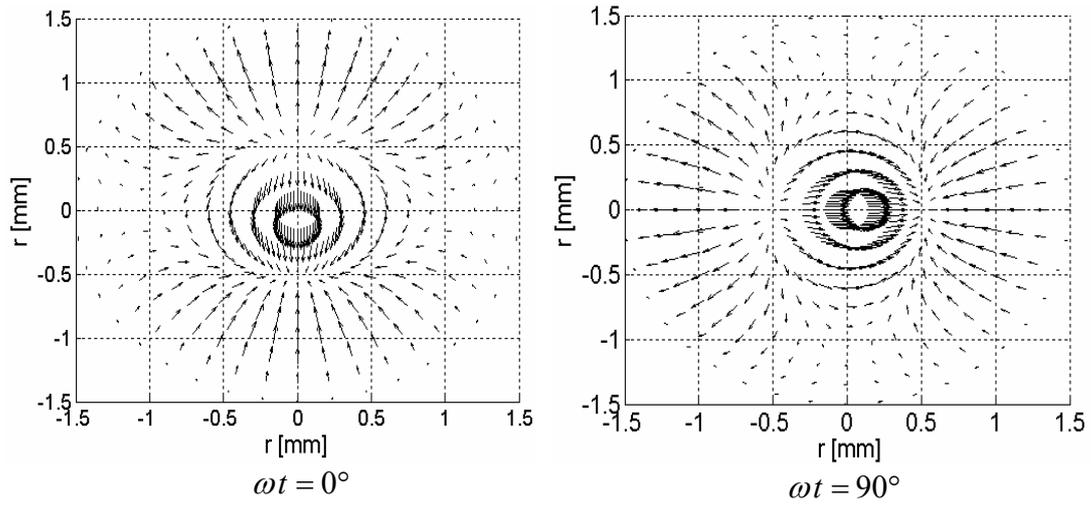

Fig. 4. A gallery of the analytically derived in-plane magnetic field distributions on the upper plane of a ferrite disk for the 2$^{nd}$ magnetic-dipolar mode ($f$ = 8.667 GHz) at some time phases. (A qualitative picture).

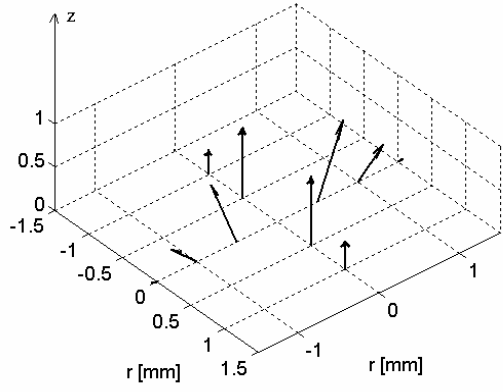
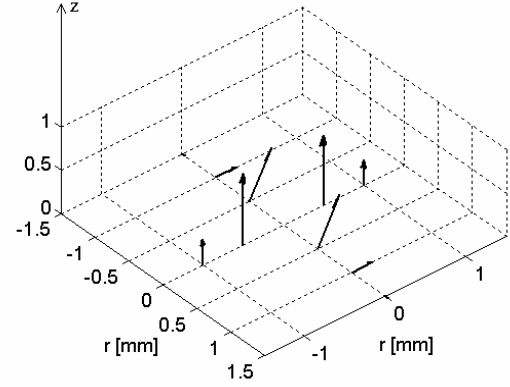

$2\omega t = 0°$          $2\omega t = 90°$

(*a*)

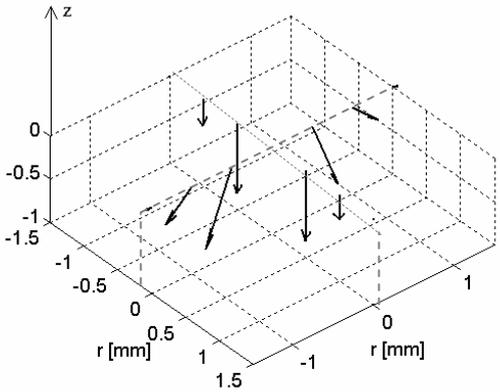
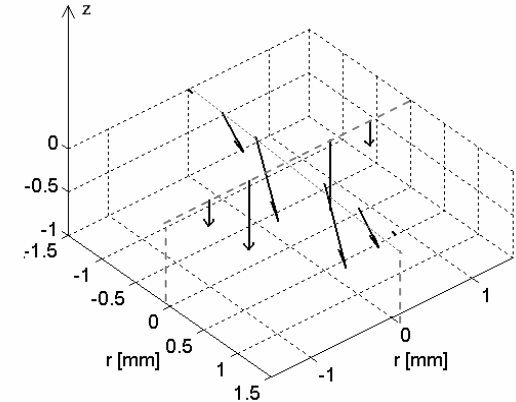

$2\omega t = 0°$          $2\omega t = 90°$

(*b*)

Fig. 5. Space orientations of vector $\vec{m} \times (\vec{\nabla} \times \vec{m})$ on the upper (a) and lower (b) planes of a ferrite disk at some time phases for the 1st MDM ($f$ = 8.548 GHz). (A qualitative picture).

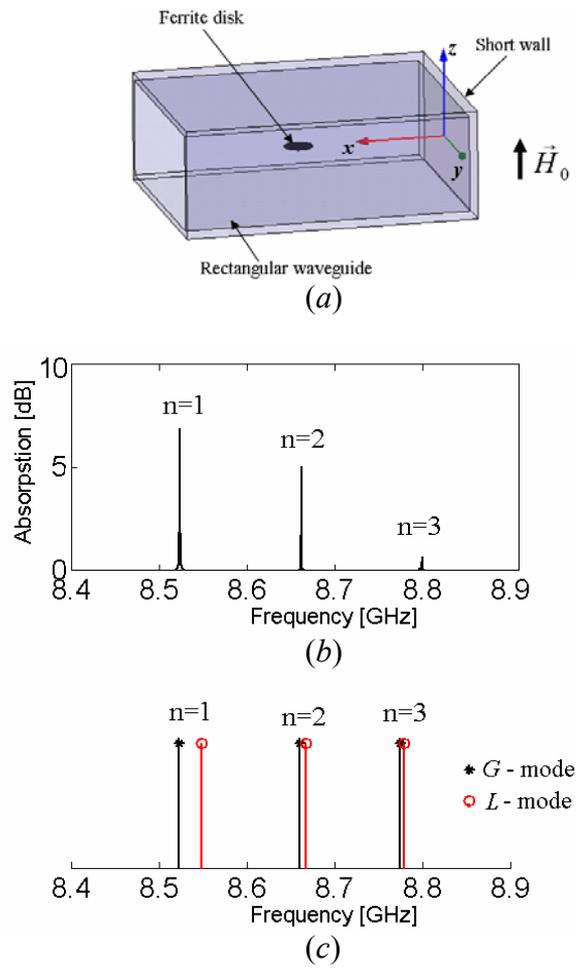

Fig. 6. Spectral characteristics for a thin ferrite disk. (a) The HFSS model of a short-wall rectangular waveguide section with a normally magnetized ferrite disk; (b) absorption peak positions obtained from the numerical simulation; (c) the calculated peak positions for the *G*-MDM and *L*-MDM (for the order parameter $\ell = +1$).

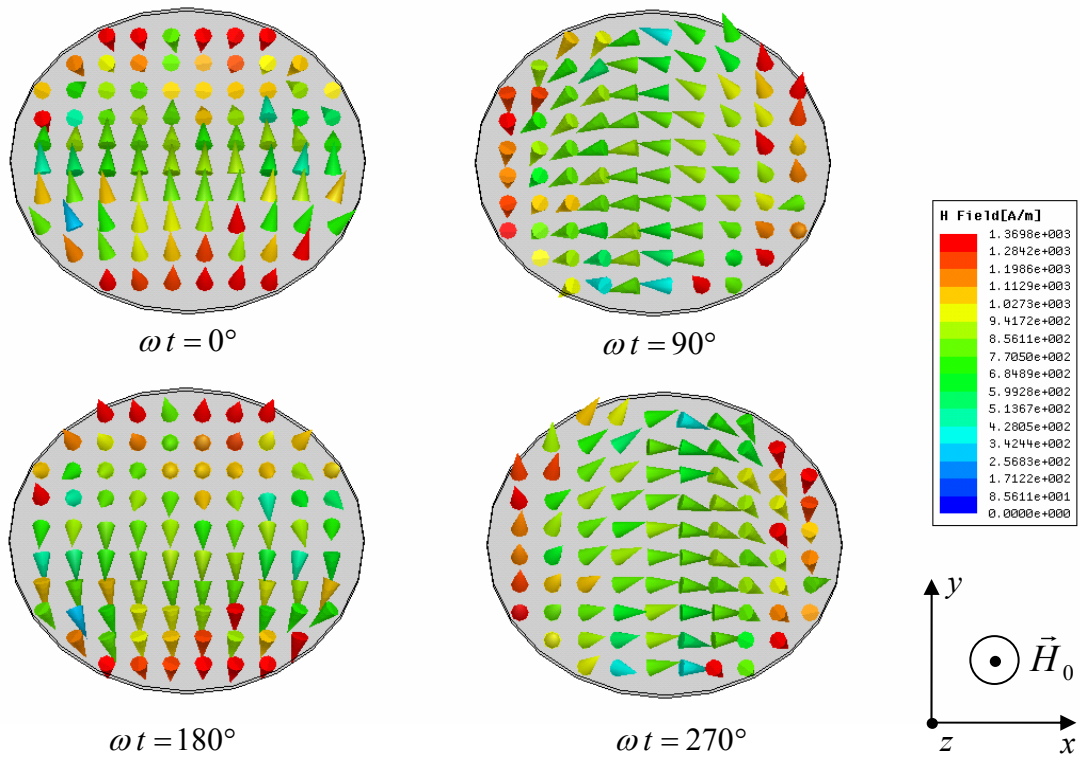

Fig. 7. A perspective view for the numerically modeled magnetic field distributions on the upper plane of a ferrite disk for the for the 1$^{st}$ resonance state ($f$ = 8.52 GHz) at different time phases.

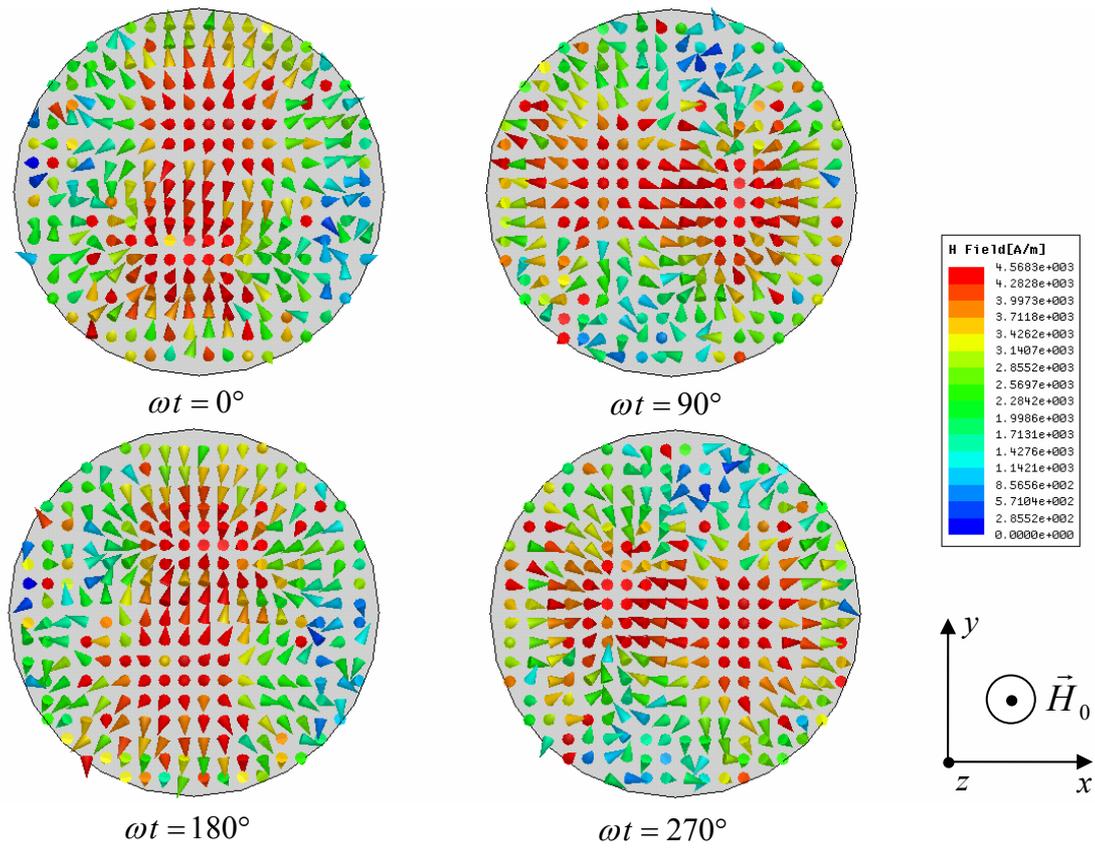

Fig. 8. A top view for the numerically modeled magnetic field distributions on the upper plane of a ferrite disk for the for the 2nd resonance state ($f$ = 8.66 GHz) at different time phases.

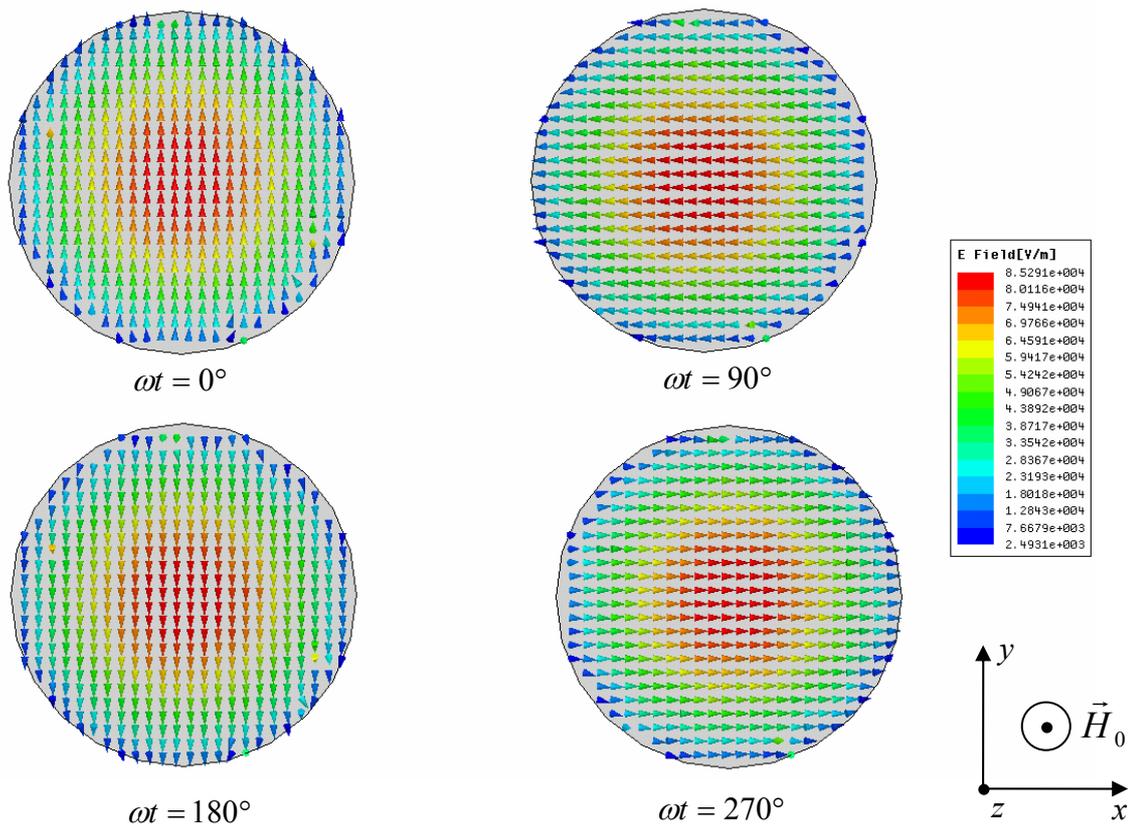

Fig. 9. A top view for the numerically modeled electric field distributions on the upper plane of a ferrite disk for the for the 1st resonance state ($f$ = 8.52 GHz) at different time phases.

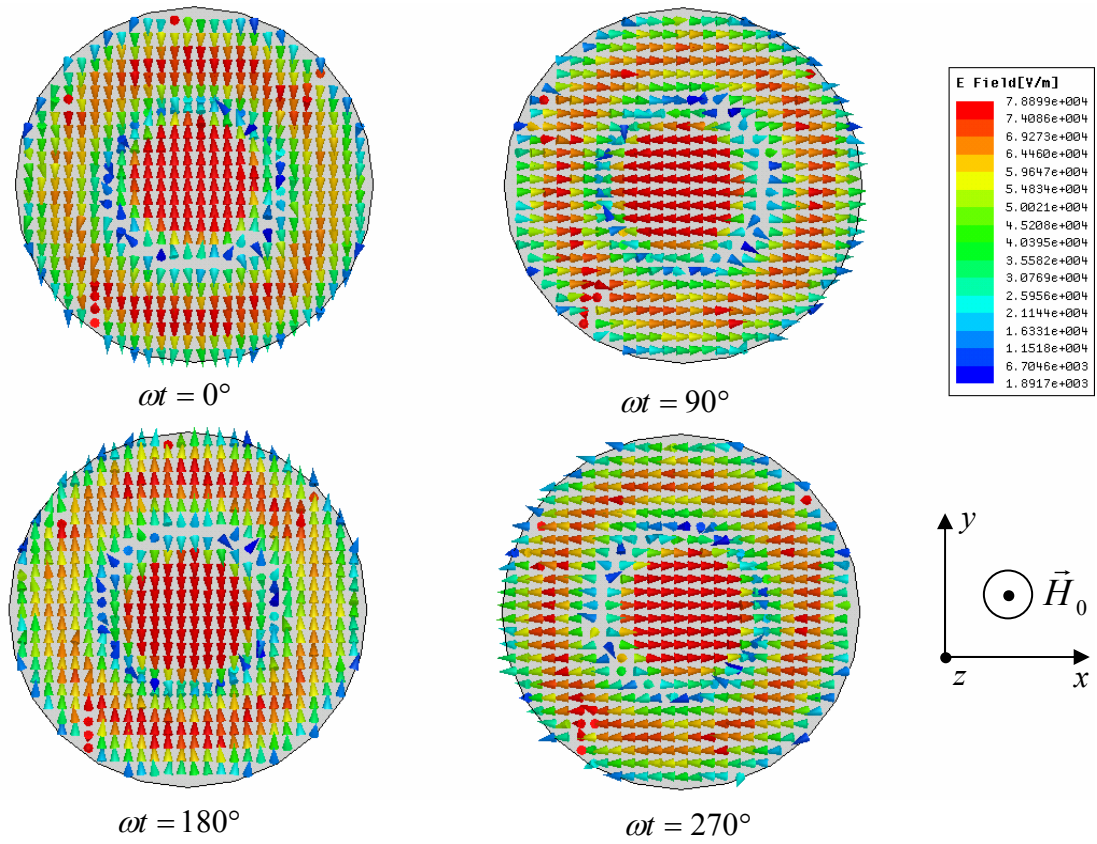

Fig. 10. A top view for the numerically modeled electric field distributions on the upper plane of a ferrite disk for the for the 2$^{nd}$ resonance state ($f$ = 8.66 GHz) at different time phases.

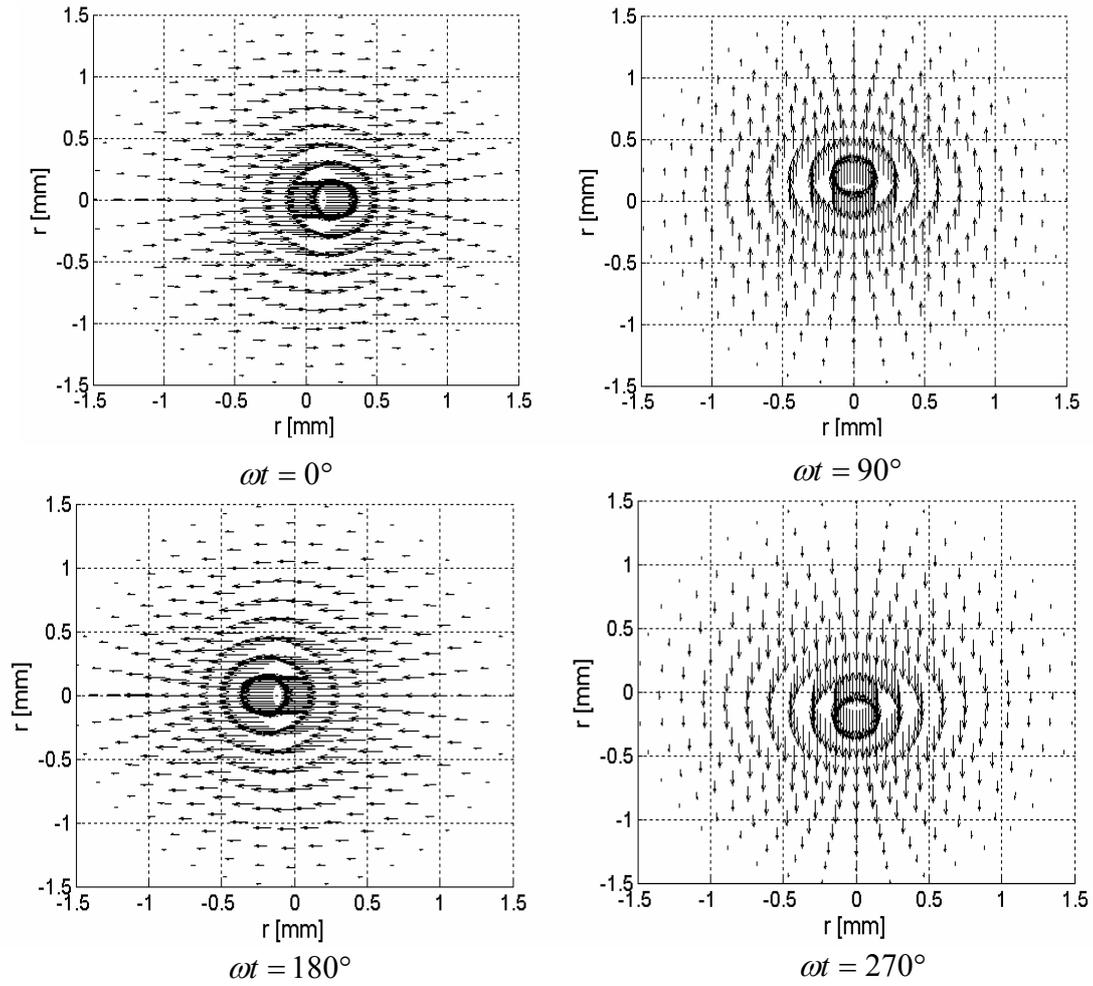

Fig. 11. A gallery of the analytically derived in-plane effective-electric-current distributions on the upper plane of a ferrite disk for the 1$^{st}$ magnetic-dipolar mode ($f$ = 8.548 GHz) at different time phases. (A qualitative picture).

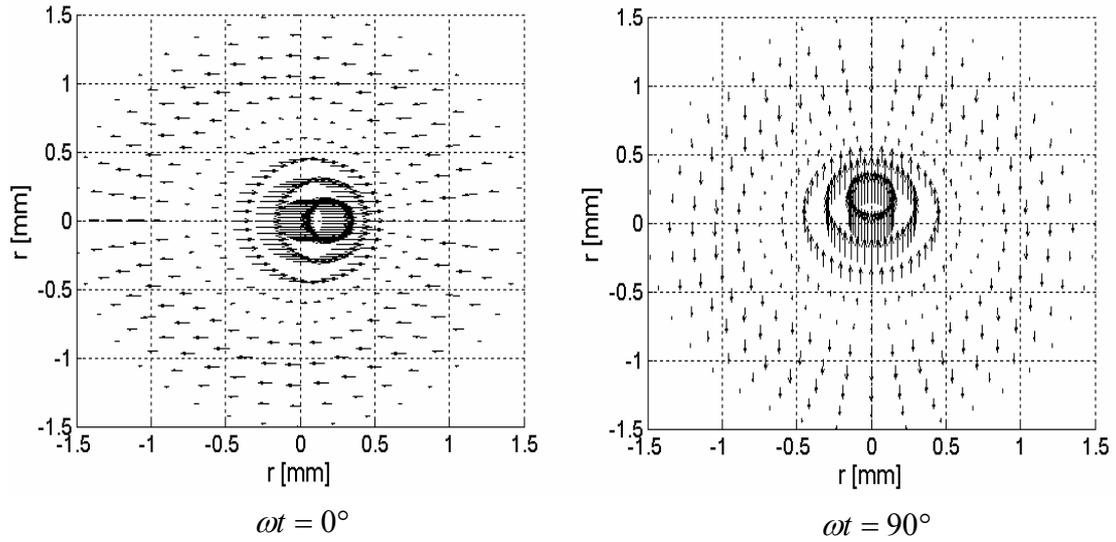

Fig. 12. A gallery of the analytically derived in-plane effective-electric-current distributions on the upper plane of a ferrite disk for the 2$^{nd}$ magnetic-dipolar mode ($f$ = 8.667 GHz) at some time phases. (A qualitative picture).

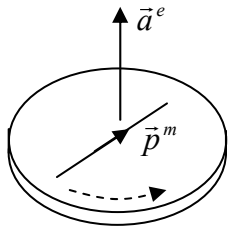

Fig. 13. The ME-particle model of a MDM ferrite disk. Since the spectra of *G*- and *L*-MDMs are almost degenerate the ME properties of a ferrite disk are observed as the *G*-mode electric (anapole) moment $\vec{a}^e$ normal to a ferrite disk together with the *L*-mode magnetic moment $\vec{p}^m$ rotating in the disk plane.

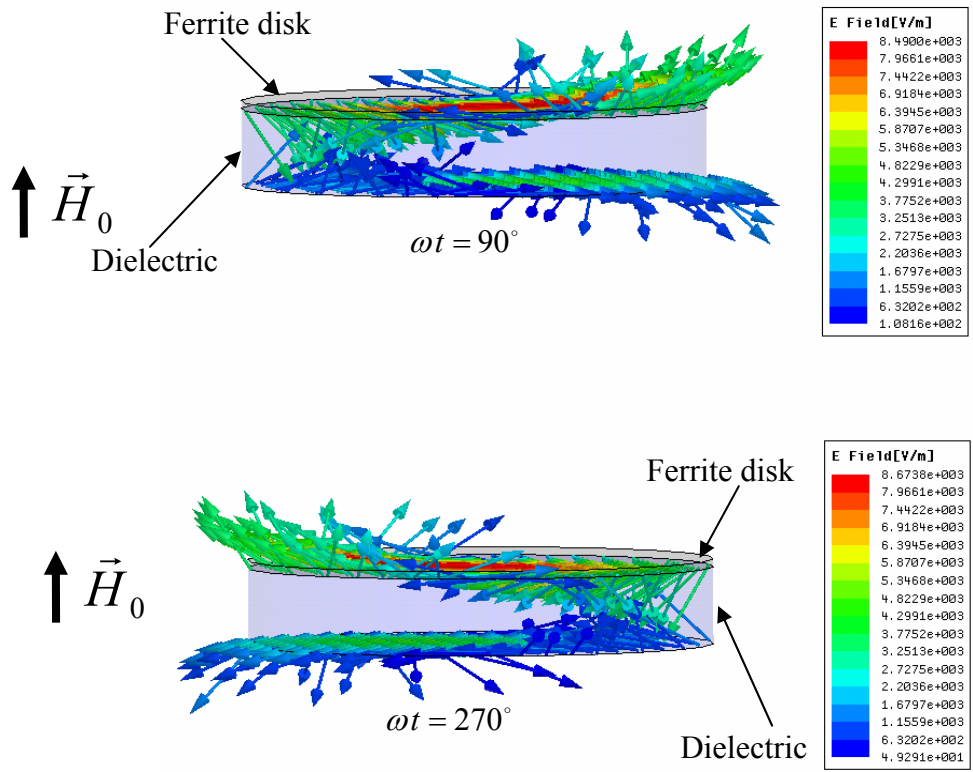

Fig. 14. The electric field distributions for the 1st resonance state for a sample composed as a thin ferrite disk loaded with a dielectric disk. One has an in-plane rotating electric dipole with topological electric charges. A rotating electric dipole is in the same orientation as an in-plane rotating magnetic dipole.

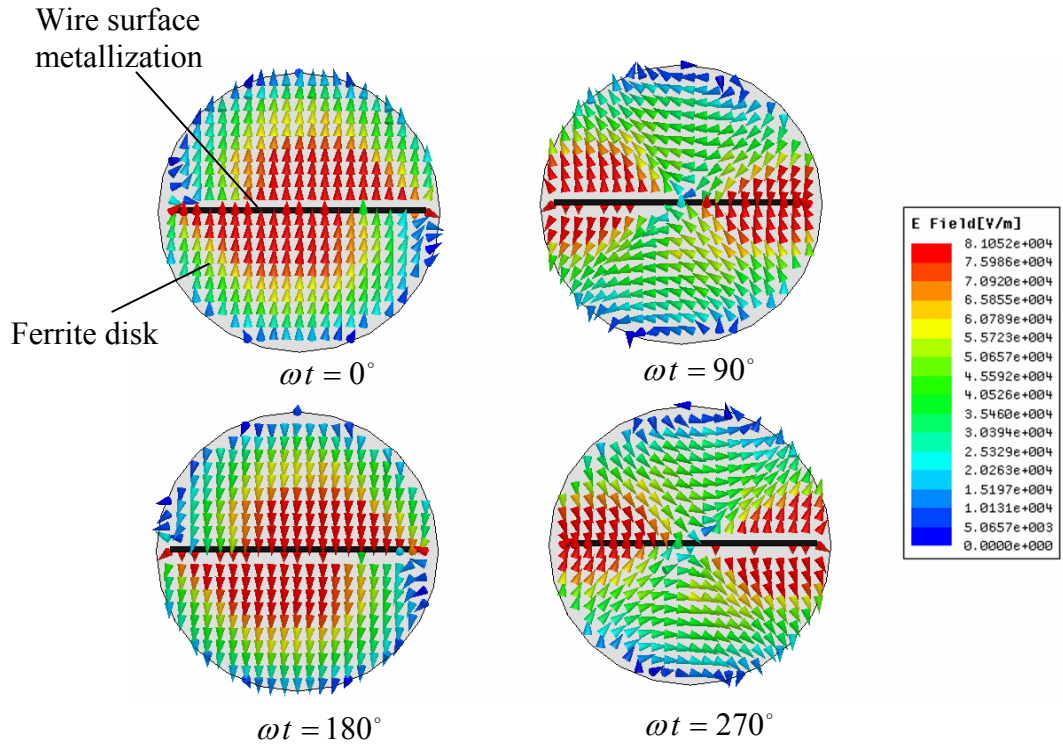

Fig. 15. The electric field distributions for the 1st resonance state for a sample composed as a thin ferrite disk with a wire surface metallization. One has an in-plane non-rotating electric dipole and an in-plane rotating magnetic dipole.